\documentstyle[11pt, aaspp4]{article} % para preprint
\begin{document}
\title {OSCILLATOR STRENGTHS AND DAMPING CONSTANTS FOR ATOMIC
LINES IN THE J AND H BANDS}

\author{\bf Jorge Mel\'endez and Beatriz Barbuy}
\affil{Universidade de S\~ao Paulo, IAG, Dept. de Astronomia,
 CP 3386, S\~ao Paulo 01060-970, Brazil \\ E-mail:
jorge@iagusp.usp.br, barbuy@orion.iagusp.usp.br}

\slugcomment{Submitted to the The Astrophysical Journal}
\slugcomment{Send proofs to:  J. Melendez}

\begin{abstract}

We have built a line list in the near-infrared J and H bands
(1.00-1.34,  1.49-1.80 $\mu$m)
by gathering a series of laboratory and computed line lists.
Oscillator strengths and damping constants were computed or
obtained by fitting  the solar spectrum.

The line list presented in this paper is, to our knowledge,
the most complete one now available, and supersedes previous
lists.
\end{abstract}

\section{Introduction}
Spectroscopy in the  near-infrared (NIR) is at present essentially
limited to molecular bands, partly due to a lack of a line
list with reliable atomic constants.

Lists of lines in the NIR are given for individual
elements and in most cases only the identification 
(element/wavelength) is provided.

With the advent of improved infrared detectors, it is possible
to carry out abundance studies using these wavelength regions.
High excitation potentials typical of the lines in these wavelength
regions, make such stellar NIR spectroscopy suitable
as a complement to the optical region, by using particular lines such
as \ion{S}{1}, \ion{Mn}{1} (lines with lower excitation potential) in giants,
or applicable to hot dwarf turn-off stars.

The NIR is otherwise the only means to obtain spectra
for objects located in reddened regions such as
the Galactic bulge, for which the optical regions are not
reachable due to a high extinction (e.g. Ramirez et al. 1998;
Sellgren et al. 1998).

The line list is presented in Sec. 2 and the derivation of the
damping constants and oscillator strengths are described in 
Sec. 3 and Sec. 4, respectively.  Concluding remarks are 
given in Sec. 5.

\section{ Line list}

The atomic line data was started with the lists by
 Ramsauer et al. (1995)  and Swensson et al. (1973), 
and progressively available line lists were added.
Most additional lines correspond to Fe and Si, but
lines of the elements  C, Mg, Ca, Ti, Cr, Mn, Ni 
were also included.
The literature used for each element is given in Table 1.
Besides line lists for individual elements, we used
the general line lists of Swensson et al. (1973), 
Bi\'emont \& Grevesse (1973), Bi\'emont (1976), Outred (1978), 
Solanki et al (1990), Livingston \& Wallace (1991), 
Wallace et al. (1993),  Kurucz (1995), Ramsauer et al. (1995), 
Hirata \& Horaguchi (1995).

We made some new identifications of solar line features surveying 
the most recent laboratory and theoretical work and using the 
telluric free Kitt Peak Solar Atlas (Livingston \& Wallace 1991, 
Wallace et al. 1993). A new identification is based on the intensities 
and on the correspondence between the observed solar  
($\lambda_{\odot}$) and laboratory ($\lambda_{lab}$) wavelengths 
(or theoretical wavelength, $\lambda_{th}$, in some cases).

We note that in the Wallace et al. (1993) solar atlas, all lines of 
\ion{Al}{1} are assigned to \ion{Al}{2}, probably due to a misprint.

\ion{Ni}{1} lines show isotopic splitting, and for cases in
which they are not severely 
blended with other lines, we included the isotopic components in the list. 
Isotopic splitting of Ni is remarkable in the NIR, and was reported 
by Brault \& Holweger (1981), Bi\'emont et al. (1986) and 
Litz\'en et al. (1993). The isotopic species ($^{58,60,62,64}$Ni) are 
indicated in column 6 of Tables 2 and 3.

Molecular lines of CN A$^2 \Pi$ - X$^2 \Sigma$ and vibration-rotation 
CO X$^1 \Sigma^+$ bands were included in the synthetic spectra
calculations.
The line lists for CN  were made available by S. P. Davis
(see {\it http://sumner.berkeley.edu}) and the CO line lists
were adopted from Goorvitch (1994).

For CN and CO the Honl-London factors were computed using
formulae by Kov\'acs (1969). Rotational Franck-Condon factors for CN
were computed following Dwivedi et al. (1978). For CO the 
expectation values of the effective dipole moment operator were taken
from Goorvitch (1994). The electronic transition moment of CN bands 
were obtained from a fit to CN unblended solar lines. 
The dissociation potentials adopted are $D_0$(CN) = 7.65 eV 
(Bauschlicher et al. 1988) and 
$D_0$(CO) = 11.09 eV (Huber \& Herzberg 1979).

\section{Damping constants}

The van der Waals collision broadening is given by
$\gamma$$_6$/N$_{\rm H}$ = 17 v$^{3/5}$ C$_6^{2/5}$
where v is velocity, N$_{\rm H}$ is the number density of hydrogen
and C$_6$ is the interaction constant for van der Waals pressure
broadening.

C$_6$ was computed by using the cross sections $\sigma$(v) given by 
Anstee \& O'Mara (1995) for atomic transitions s-p, p-s,
Barklem \& O'Mara (1997) for transitions p-d, d-p
and Barklem, O'Mara \& Ross (1998) appliable to transitions involving
excited states d-f, f-d.

These tables give $\sigma$(v) in terms of the principal effective
quantum numbers (n$^*$) of the upper and lower states, 
n$^*$ = 1 /$ \sqrt {(\chi_{ion}(E)-\chi_{exc}(E))/ \chi_{ion}(H)}$,
where $\chi_{ion}(H)$ and $\chi_{ion}(E)$ are 
the ionization energies of hydrogen and the element $E$ respectively, 
and $\chi_{exc}(E)$ is the excitation potential. 

Because the tables of cross sections do not cover very high excitation 
lines (high n$^*$ numbers) we had to make extrapolations 
(until $\Delta$ n$^* \approx 0.4$). In those cases the extrapolation
is indicated by a letter ``E'' next to the C$_6$ values given 
in Tables 2 and 3.

In cases where it was not possible to use the tables of cross section, 
we computed the C$_6$ values by employing the classical formula: 
$C_6 \; = \; 6.46 \times 10^{-34} \bar{\Delta r^2}$, using the 
Coulomb approximation to calculate the mean square radius 
$\bar{r^2}$ of the upper and lower states. We used an enhancement 
factor $E$ of 2.5, which means a 
$C_6$ value 10 times larger than the classical van der Waals value 
($C_6 \; \propto \; \gamma_6^{2.5}$). The lines for which this method
was used are indicated with a letter ``M'' next to the C$_6$ values.

Finally, in some cases we obtained the C$_6$ values from a fit to 
the solar spectrum. In those cases a letter ``A'' is given next to
the C$_6$ values.

\section{Astrophysical oscillator strengths}

The oscillator strengths are derived from an iterative fit of
synthetic spectra to the
solar spectum (Livingston \& Wallace 1991, Wallace et al. 1993).
 The code for spectrum synthesis is described in 
Barbuy (1981, 1982). The synthetic spectra are calculated with the LTE 
approximation. The adopted solar abundances are those reported by 
Grevesse et al. (1996).

Some strong lines with large central depths show NLTE effects and the
$gf$ values are probably overestimated. We tried to avoid it
by fitting the wings.

The errors in the oscillator strengths are indicated with a superscript
to the $gf$ value in Tables 2 and 3, by $a$: $\leq$ 0.05 dex, 
$b$: $\leq$ 0.10 dex, $c$: $\leq$ 0.15 dex, $d$: $>$ 0.15 dex.
The quoted errors are estimated basically by uncertainties in
fixing the continuum, blends, NLTE, or  solar line profiles
showing residuals of blends with heavy telluric absorption.
The letter {\it f} is used to indicate that the $gf$ value of the
line was fixed while the oscillator strengths of the 
neighbouring lines
were fitted; in these cases, the source of the $gf$ fixed value is
in most cases Kurucz (1995).

\section{Conclusions}

We present a complete line list of atomic lines,
typical of late A-F-G-K stars, 
in the J and H bands. 
 The only such work available
in the literature is the one by Stalin et al. (1997),
which comprises however only about 15 \% of the lines 
contained in our list.

A major contribution provided in the present paper is the computation of
accurate damping constants C$_6$ line-by-line, 
and subsequent derivation of $gf$ values,
obtained by fitting the solar spectrum, 
 resulting in accurate oscillator strengths.

\acknowledgements 
We acknowledge partial
financial support from CNPq and FAPESP.
J.M. acknowledges the FAPESP PhD fellowship n$^o$ 97/0109-8. 
NSO/Kitt Peak FTS data used here were produced by NSF/NOAO.

\begin{deluxetable}{rlcccc}
\tablenum{1}
\tablewidth{0pt}
\footnotesize
\tablecaption{Summary of Atomic Lines}
\tablehead{
\colhead{Z}  &
\colhead{Element}      &
\colhead{Abundance}      &
\colhead{n$^{\rm o}$ lines J/H}      &
\colhead{wavelength range (\AA)} &
\colhead{ref.} }
\startdata
1  & H I   & 12.00 & 3/5   &  10049 - 12818, 15261 - 17362 & [1] \nl
2  & He I  & 10.99 & 3/0   &  10829 - 10830                & [2] \nl
6  & C I   & 8.55  & 58/83 & 10124 - 12897, 15050 - 17966  & [3,4,5,6] \nl
7  & N I   & 7.97  & 11/0  &  10105 - 10758                & [7]  \nl
8  & O I   & 8.87  & 5/0   &  11297 - 13165                & [8,9,10,11] \nl
11 & Na I  & 6.33  & 16/2  &  10182 - 12679, 16374 - 16389 & [12,13] \nl
12 & Mg I  & 7.58  & 28/51 &  10299 - 12433, 15025 - 17762 & [14,15,16] \nl
12 & Mg II & 7.58  & 6/0   &  10092 - 10952                & [16,17] \nl
13 & Al I  & 6.47  & 7/5   &  10768 - 13151, 16719 - 17708 & [18,19] \nl
14 & Si I  & 7.55  & 131/50 & 10001 - 13326, 15244 - 17623 & [20,21,22] \nl
15 & P I   & 5.45  & 10/4  &  10084 - 11183, 15711 - 17112 & [23,24]  \nl
16 & S I   & 7.21  & 14/16 &  10455 - 11602, 15400 - 16597 & [25,26] \nl
19 & K I   & 5.12  & 8/3   &  11020 - 12522, 15163 - 15618 & [12,13] \nl
20 & Ca I  & 6.36  & 45/7  &  10249 - 13318, 15067 - 16197 & [27] \nl
20 & Ca II & 6.36  & 5/2   &  11429 - 11950, 16561 - 16650 & [27,28] \nl
21 & Sc I  & 3.17  & 0/1   &    17523                      & [29] \nl
22 & Ti I  & 5.02  & 56/26 &  10003 - 13077, 15016 - 17447 & [30] \nl
22 & Ti II & 5.02  & 1/0   &      10691                    & [31] \nl
23 & V I   & 4.00  & 5/0   &  10230 - 11195                & [32] \nl
24 & Cr I  & 5.67  & 40/14 &  10080 - 13217, 15069 - 17938 & [33,34] \nl
25 & Mn I  & 5.39  & 73/62 &  12899 - 13319, 15159 - 17746 & [35,36,37] \nl
26 & Fe I  & 7.50  & 363/800 & 10007 - 13392, 14904 - 17986 & [38...46]\nl
26 & Fe II & 7.50  & 13/0 &  10173 - 12264           & [43,47,48,49,50] \nl
27 & Co I  & 4.92  & 12/5 &  10019 - 12880, 16130 - 16997   & [51,52] \nl
28 & Ni I  & 6.25  & 51/109 &  10049 - 13264, 15032 - 17988 & [53,54] \nl
30 & Zn I  & 4.60  & 3/0 &  11054 - 13197                  & [55] \nl
32 & Ge I  & 3.41  & 7/0 &  10039 - 11715                  & [56] \nl
38 & Sr II & 2.90  & 3 /0 &  10037 - 10915                 & [56] \nl
39 & Y II  & 2.24  & 5/0 &  10105 - 10605                  & [57]\nl
57 & La II & 1.22  & 1/0 &    11874                        & [56] \nl
63 & Eu II & 0.51  & 1/0 &    10020                        & [56] \nl

\enddata
\tablerefs{
[1] Garcia \& Mack (1965),
[2] Martin (1960),
[3] Johansson \& Litz\'en (1965),
[4] Johansson (1965),
[5] Bi\'emont (1973),
[6] Bi\'emont et al. (1993),
[7] Eriksson (1958),
[8] Eriksson \& Isberg (1963b),
[9] Chang et al. (1988),
[10] Bi\'emont et al. (1991),
[11] Hibbert et al. (1991),
[12] Risberg (1956),
[13] Johansson (1961),
[14] Risberg (1965),
[15] Swensson \& Risberg (1966),
[16] Bi\'emont \& Brault (1986),
[17] Risberg (1955),
[18] Eriksson \& Isberg (1963a),
[19] Bi\'emont \& Brault (1987),
[20] Litzen (1964), 
[21] Radziemski \& Andrew (1965),
[22]Lambert \& Warner (1968),
[23] Svendenius (1980),
[24] Bi\'emont et al. (1994),
[25] Jakobsson (1967),
[26] Bi\'emont et al. (1993),
[27] Risberg (1968),
[28] Edl\'en \& Risberg (1956),
[29] Kaufman \& Sugar (1988),
[30] Forsberg (1991),
[31] Huldt et al. (1982),
[32] Palmeri et al. (1997),
[33] Kiess (1953),
[34] Bi\'emont et al. (1985b),
[35] Swensson (1965),
[36] Taklif (1990),
[37] Mel\'endez (1999),
[38] Litz\'en (1976),
[39] Litz\'en \& Verg\`es (1976),
[40] Bi\'emont et al. (1985a), 
[41] Johansson \& Learner (1990),
[42] O'Brian et al. (1991),
[43] Nave et al. (1992),
[44] Nave \& Johansson (1993a),
[45] Nave \& Johansson (1993b),
[46] Nave et al. (1994),
[47] Johansson (1977),
[48] Johansson (1978),
[49] Rosberg \& Johansson (1992),
[50] Rosberg et al. (1993),
[51] Russell et al. (1940),
[52] Pickering \& Thorne (1996),
[53] Bi\'emont et al. (1986),
[54] Litz\'en et al. (1993),
[55] Johansson \& Contreras (1968),
[56] Swensson et al. (1973),
[57] Nilsson et al. (1991). }
 
\end{deluxetable}


\begin{references}

\reference{} Anstee, S. D., O'Mara, B. J. 1995, MNRAS, 276, 859.

\reference{} Barbuy, B. 1981, A\&AS, 101, 365-368

\reference{} Barbuy, B. 1982, {\it Ph.D. thesis}, Univ. Paris VII

\reference{} Barklem, P. S., \& O'Mara, B. J. 1997,  MNRAS, 290, 102

\reference{} Barklem, P. S., O'Mara, B. J., \& Ross, J. E. 1998, 
MNRAS, 296, 1057

\reference{} Bauschlicher, C. W., Langhoff, S. R., 
\& Taylor, P. R. 1988, ApJ, 332, 531

\reference{} Bi\'emont, E. 1973, Solar Physics, 32, 117

\reference{} Bi\'emont, E. 1976, A\&AS, 26, 89

\reference{} Bi\'emont, E., Brault, J. W., Delbouille, L., 
\& Roland, G. 1985a, A\&AS, 61, 107

\reference{} Bi\'emont, E., Brault, J. W., Delbouille, L., 
\& Roland, G. 1985b, A\&AS, 61, 185

\reference{} Bi\'emont, E., Brault, J. W., Delbouille, L., 
\& Roland, G. 1986, A\&AS, 65, 21

\reference{} Bi\'emont, E., \& Brault, J. W. 1986, Physica Scripta, 34, 751

\reference{} Bi\'emont, E., \& Brault, J. W. 1987, Physica Scripta, 35, 286

\reference{} Bi\'emont, E., \& Grevesse, N. 1973, Atomic Data and 
Nuclear Data Tables, 12, 217

\reference{} Bi\'emont, E., Hibbert, A., Godefroid, M., Vaeck, N., 
\& Fawcett, B. C. 1991, ApJ, 375, 818

\reference{} Bi\'emont, E., Martin, F., Quinet, P., \& Zeippen, C. J. 
1994, A\&A, 283, 339

\reference{} Bi\'emont, E., Quinet, P., \& Zeippen, C. J. 1993, A\&AS, 102, 435

\reference{} Brault, J. W., \& Holweger, H. 1981, ApJ, 249, L43

\reference{} Chang, E. S., Barowy, W. M., \& Sakai, H. 1988,  
Physica Scripta, 38, 22

\reference{} Dwivedi, P. H., Branch, D., Huffaker, J. N.,
\& Bell, R. A.  1978, ApJS, 36, 573

\reference{} Edl\'en, B. 1953, J. Opt. Soc. Amer., 43, 339

\reference{} Edl\'en, B. 1966, Metrologia, 2, 71

\reference{} Edl\'en, B., \& Risberg, P. 1956, Ark. Fysik, 10, 553

\reference{} Eriksson, K. B. S. 1958, Ark. Fysik, 13, 429

\reference{} Eriksson, K. B. S., \& Isberg, H. B. S. 1963a, 
Ark. Fysik, 23, 527

\reference{} Eriksson, K. B. S. \& Isberg, H. B. S. 1963b, 
Ark. Fysik, 24, 549

\reference{} Forsberg, P. 1991,  Physica Scripta, 44, 446

\reference{} Garcia, J. D., \& Mack, J. E. 1965, J. Opt. Soc. Amer., 55, 654

\reference{} Goorvitch, D. 1994, ApJS, 95, 535

\reference{} Grevesse, N., Noels, A., 
        Sauval, J. 1996, in ASP Conf.  Ser.  99, eds.  S.S. Holt, G. 
        Sonneborn, p.  117 

\reference{} Hirata, R., \&  Horaguchi, T. 1995, 
Atomic spectral line list, Kyoto University.

\reference{} Hibbert, A., Bi\'emont, E., Godefroid, M., \& Vaeck, N. 1991, 
J. Phys. B: At. Mol. Opt. Phys., 24, 3943

\reference{} Huber, K. P., \& Herzberg, G. 1979, Constants of Diatomic
Molecules, Van Nostrand Reinhold, New York.

\reference{} Huldt, S., Johansson, S., Litz\'en, U., \& Wyart, J. F. 1982,
Physica Scripta, 401

\reference{} Jakobsson, L. R. 1967, Ark. Fysik, 34, 19

\reference{} Johansson, I. 1961, Ark. Fysik, 20, 135

\reference{} Johansson, I., \& Contreras, R. 1968, Ark. Fysik, 37, 513

\reference{} Johansson, L. 1965, Ark. Fysik, 31, 201

\reference{} Johansson, L., \& Litz\'en, U. 1965, Ark. Fysik, 29, 175

\reference{} Johansson, S. 1977, MNRAS, 178, 17P

\reference{} Johansson, S. 1978,  Physica Scripta, 18, 217

\reference{} Johansson, S., \& Learner, R. C. M. 1990, ApJ, 354, 755

\reference{} Kaufman, V., \& Sugar, J. 1988, J. Phys. Chem. Ref. Data, 17, 1679

\reference{} Kiess, C. C. 1953,  J. Res. National Bureau of Standards, 51, 247

\reference{} Kov\'acs, I. 1969, Rotational Structure in the Spectra of 
Diatomic Molecules, American Elsevier Publishing Company, New York

\reference{} Kurucz, R. 1995, CD-ROM 23, Harvard-Smithsonian Center for Astrophysics.

\reference{} Lambert, D. L., \& Warner, B. 1968, MNRAS, 139, 35

\reference{} Litz\'en, U. 1964, Ark. Fysik, 28, 239

\reference{} Litz\'en, U. 1976, Physica Scripta, 14, 165

\reference{} Litz\'en, U., Brault, J. W., \& Thorne, A.P. 1993, 
Physica Scripta, 47, 628

\reference{} Litz\'en, U., \& Verg\`es, J. 1976, Physica Scripta, 13, 240

\reference{}  Livingston, W., \& Wallace, L. 1991, An Atlas of the 
Solar Spectrum in the Infrared (1.1 to 5.4 $\mu$m), 
National Solar Obs., Tech. Rep. 91-001, Tucson, AZ

\reference{} Martin, W. C. 1960, J. Res. National Bureau of Standards, A, 64, 19

\reference{} Mel\'endez, J. 1999, submitted to MNRAS

\reference{} Nadeau, D. 1988, ApJ, 325, 480

\reference{} Nave, G., \& Johansson, S. 1993a, A\&A, 274, 961

\reference{} Nave, G., \& Johansson, S. 1993b, A\&AS, 102, 269

\reference{} Nave, G., Johansson, S., Learner, R. C. M., Thorne, P., 
\& Brault, J. W. 1994, ApJS, 94, 221

\reference{} Nave, G., Learner, R. C. M., Murray, J. E., Thorne, A. P., 
\& Brault, J. W. 1992, Journal de Physique II France 2, 913

\reference{} Nilsson, A. E., Johansson, S., \& Kurucz, R. L. 1991, 
Physica Scripta, 44, 226

\reference{} O'Brian, T. R., Wickliffe, M. E., Lawler, J. E., Whaling, W.,
Brault, J. W. 1991, J. Opt. Soc. Amer., B8, 1185

%\reference{} Origlia, L., Moorwood, A.F.M., Oliva, E.: 1993, A\&A, 280, 536.

\reference{} Outred, M. 1978, J. Phys. Chem. Ref. Data, 7, 1

\reference{} Palmieri, P., Bi\'emont, E., Quinet, P., Dembcz\'ynski, J.,
Szawiola, G., \& Kurucz, R. L. 1997, Physica Scripta, 55, 586

\reference{} Pickering, J. C., \& Thorne, A. P. 1996, ApJS, 107, 761

\reference{} Radziemski, L. J., \& Andrew, K. L. 1965, 
J. Opt. Soc. Amer., 55, 474

\reference{} Ramirez, S.V., Sellgren, K., Terndrup, D.M.,
Carr, J.S., Balachandran, S., Blum, R.D. 1998, in IAU Symp. 184,
eds. Y. Sofue, p. 67

\reference{} Ramsauer, J., Solanki, S. K., \& Bi\'emont, E. 1995, A\&AS, 113, 71

\reference{} Risberg, G. 1965, Ark. Fysik, 28, 381

\reference{} Risberg, G. 1968, Ark. Fysik, 37, 231

\reference{} Risberg, P. 1955, Ark. Fysik, 9, 483

\reference{} Risberg, P. 1956, Ark. Fysik, 10, 583

\reference{} Rosberg, M., \& Johansson, S. 1992, Physica Scripta, 45, 590

\reference{} Rosberg, M., Litzen, U., \& Johansson, S. 1993, MNRAS, 262, L1

\reference{} Russell, H. N., King, R. B., \& Moore, C. E. 1940, 
Phys. Rev., 58, 407


\reference{} Sellgren, K., 
Carr, J.S., Balachandran, S. 1998, in IAU Symp. 184,
eds. Y. Sofue, p. 21

\reference{} Solanki, S. K., Bi\'emont, E., \& M\"urset, U. 1990, A\&AS, 83, 307

\reference{} Stalin, C. S., Trivedi, C., Sihna, K., \& Sanwal, B. B. 1997,
Bull. Astr. Soc. India, 25, 353

\reference{} Svendenius, N. 1980, Physica Scripta, 22, 240

\reference{} Swensson, J.W. 1965, Ark. Fisik, 32, 463

\reference{} Swensson, J.W., Benedict, W. S., Delbouille, L., \& Roland, G. 
1973, M\'em. Soc. R. Sci. Li\`ege, vol. 5

\reference{} Swensson, J. W., \& Risberg, G. 1966, Ark. Fisik, 31, 237

\reference{} Taklif, A. G. 1990, Physica Scripta, 42, 69

\reference{} Wallace, L., Hinkle, K., \& Livingston, W. 1993, 
An Atlas of the Photospheric Spectrum from 8900 to 13600 cm$^{-1}$ 
(7350 to 11230 \AA), National Solar Obs., Tech. Rep. 93-001, Tucson, AZ.


\end{references}
\end{document}